\documentclass{PoS}

\usepackage{aas_macros}
\usepackage{graphicx}
\usepackage{float}
\usepackage{url}
\usepackage{amsmath}
\usepackage{amssymb}
\usepackage{tabulary}

\title{VERITAS detection of VHE emission from the optically bright quasar OJ~287}

\ShortTitle{VERITAS detection of OJ~287}

\author{\speaker{Stephan O'Brien} for the VERITAS Collaboration \thanks{veritas.sao.arizona.edu} \\
        University College Dublin\\
        E-mail: \email{stephan.o-brien@ucdconnect.ie}}

\abstract{We report on the VERITAS detection of very-high-energy (VHE, $E > 100~\mathrm{GeV}$) $\gamma$-ray emission from the optically bright quasar OJ~287 which is located at a redshift of z = 0.306.
	 OJ~287 has been observed to display regular optical outbursts with a period of approximately 12 years, with the last major optical outburst having occurred in 2015.
	 To explain this periodicity, models involving a binary supermassive black hole system at the core of OJ~287, or a helical jet, have been developed.
	 Motivated by elevated {\it Swift}-XRT count rates, VERITAS observed OJ~287 in February 2017, and detected the object at >5 standard deviations above background.
	 This detection prompted further VERITAS, {\it Swift}-XRT and multiwavelength observations of the object. 
	 The results of the VERITAS observational campaign are presented. }

\FullConference{The 35th International Cosmic Ray Conference,\\
                12th July - 20th July, 2017 \\
                Busan, South Korea}

\begin{document}

\section{Introduction}

OJ~287 (R.A.: 08h 54$'$ 48.8749$''$, Dec: +20h 06$'$ 30.641$''$ (J2000) \cite{RadioLocation}), is an optically bright quasar located at a redshift of $z=0.306$ \cite{OJ_287_Redshift}.
Archival optical observations dating back to 1890 \cite{1988ApJ...325..628S} have revealed a $\sim$ 12 year outburst cycle during which 2 optical outbursts are observed separated by about 1 year.
To explain this quasi-periodic behaviour and deviations from the 12 year periodicity (see \cite{Valtonen2011b}), models consisting of a binary super-massive black hole (SMBH) system at the core of the AGN have been invoked. 
Using a primary and secondary black hole mass of $1.8\times10^{10} \mathrm{M_{\odot}}$ and $1.3\times10^{8} \mathrm{M_{\odot}}$ respectively, \cite{Valtonen2011b} accurately predicted the 2015 optical outburst, with the next predicted optical outburst expected to occur in 2019 \cite{Valtonen2016}.

\par Strong high-energy (HE) $\gamma$-ray flaring has also been observed by the {\it Fermi}-LAT occurring coincidently with other multiwavelength (MWL) activity.
A 2009 high-energy $\gamma$-ray flare showed strong correlation with flaring activity in the millimetre band suggesting that the location of the emission region to be $\sim~9~\mathrm{pc}$ from the central engine \cite{2009HEFlare_MM}.
SED modelling of this flare requires a combination of synchrotron self-Compton (SSC) and external-Compton (EC) emission models to explain the broadband emission, with the high-energy component being produced by EC emission buried within a dusty region with a temperature of $\sim 250~\mathrm{K}$, possibly illuminated by the accretion disk.
In the context of this model, VHE emission would be below the sensitivity of current-generation ground based $\gamma$-ray detectors.

\par OJ~287 was previously observed by the VERITAS Collaboration during the anticipated 2007-2008 optically active phase (December 4 2007 - January 1 2008) and additionally during 2010-2011 \cite{VERITAS-UL}. 
These observations resulted in non-detections and a 99\% upper limit on the integral flux ($E > 182~\mathrm{GeV}$) of 2.6\% of the Crab Nebula flux was obtained.
OJ~287 was also observed by the MAGIC Collaboration as part of a multiwavelength campaign (see \cite{Magic_MWL}) during the anticipated 2007 optical outburst phase between April 10th and April 13th  and between November 7th and November 9th resulting in non-detection and 95\% upper limits on the integral flux of ($E > 145~\mathrm{GeV}$) 3.3\% and ($E > 150~\mathrm{GeV}$) 1.7\% of the Crab Nebula flux, respectively.

\par In these proceedings we report on observations taken by VERITAS between December 2016 and March 2017, resulting in the detection of VHE emission spatially coincident with OJ~287 and temporally coincident with enhanced MWL activity.
The layout of these proceedings is as follows. 
In Section~\ref{sec:VERITAS} the VERITAS instrument, observations and results are discussed. 
In Section~\ref{sec:MWL} multiwavelength observations taken with the {\it Fermi} and {\it Swift} space telescopes are discussed and results are presented.
In Section~\ref{sec:Discus} the multiwavelength properties of OJ~287 are discussed and correlations between the different wavelengths are considered.
Finally in Section~\ref{sec:Summary} the key science results of this study are summarised and future plans are discussed.

\section{VERITAS Observations}
\label{sec:VERITAS} 

VERITAS is an array of four 12-m imaging atmospheric-Cherenkov telescopes (IACTs) located at the Fred Lawrence Whipple Observatory (FLWO) in southern Arizona (31$^\circ$ 40$^\prime$N, 110$^\circ$ 57$^\prime$W,  1.3km a.s.l.).
Each telescope has a camera consisting of 499 photo-multiplier tubes that provide a field of view of approximately $3.5^\circ$.
VERITAS can accurately reconstruct $\gamma$-rays with energies between $85~\mathrm{GeV}$ and $> 30~\mathrm{TeV}$, with an energy resolution of 15-25\% and an effective collection area for a $1~\mathrm{TeV}$ photon on the order of $10^5~\mathrm{m^2}$.
It has the sensitivity to detect a source with flux equal to 1\% of the Crab Nebula flux at 5 standard deviations above background ($\sigma$) in $\sim$25 hours (see \cite{VERITAS_Performance} and references therein for full details).

\par As part of its long-term monitoring program, VERITAS regularly monitors a number of multiwavelength resources.
One such resource is the {\it Swift}-XRT $\gamma$-ray ``sources of interest'' monitoring page \cite{Swift_monitoring}.\footnote{\url{http://www.swift.psu.edu/monitoring/}}
In late 2016 {\it Swift} observed OJ~287 to undergo a period of enhanced X-ray and multiwavelength activity (for example see \cite{2016ATel.9629....1G}).  
In response to this VERITAS observed OJ~287 on several occasions in December 2016 and January 2017 resulting in a non-detection. 
Due to historic X-ray count rates observed \cite{swift_historic_flare} in the {\it Swift}-XRT monitoring site, VERITAS initiated target of opportunity (ToO) observations of OJ~287 between February 1st and February 4th 2017.
This resulted in the detection of VHE emission at $>5\sigma$ and the release of an Astronomer's Telegram \cite{detection_atel}, and triggered further {\it Swift} observations of the source.
Intense VERITAS follow-up observations were taken between February 16 and March 30 2017, with many observations taken simultaneously with \textit{Swift}.

\par A total of 50 hours of post-quality cuts and dead-time adjusted data were analysed using the two standard VERITAS analysis packages (see \cite{veritas_analysis} for a detailed discussion of VERITAS analysis techniques).
Excellent agreement was found between the two packages.
Boosted decision tree (BDT) cuts \cite{bdt_cuts} optimised for soft-spectrum source analysis were used resulting in the detection of 3178 on-source and 15734 off-source events with an average on/off normalisation of 0.167 giving a total excess significance of $9.7\sigma$ \cite{liandma}. 

\begin{figure*}[!htbp]
\centering
\begin{minipage}{.5\textwidth}
  \includegraphics[width=\linewidth]{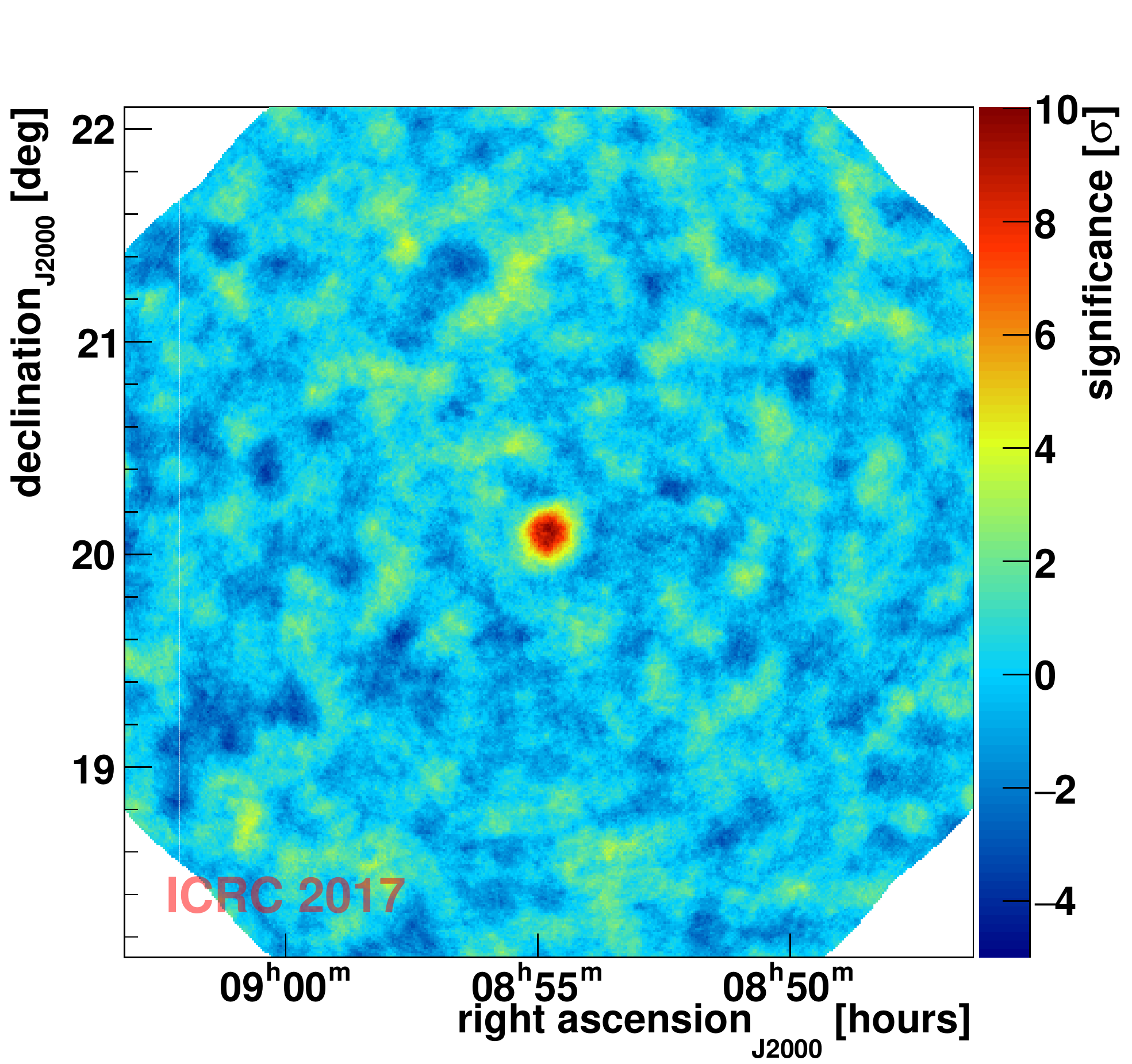}
  \caption{Significance sky map of OJ~287.}
  \label{veritas:skymap}
\end{minipage}%
\begin{minipage}{.5\textwidth}
  \includegraphics[width=\linewidth]{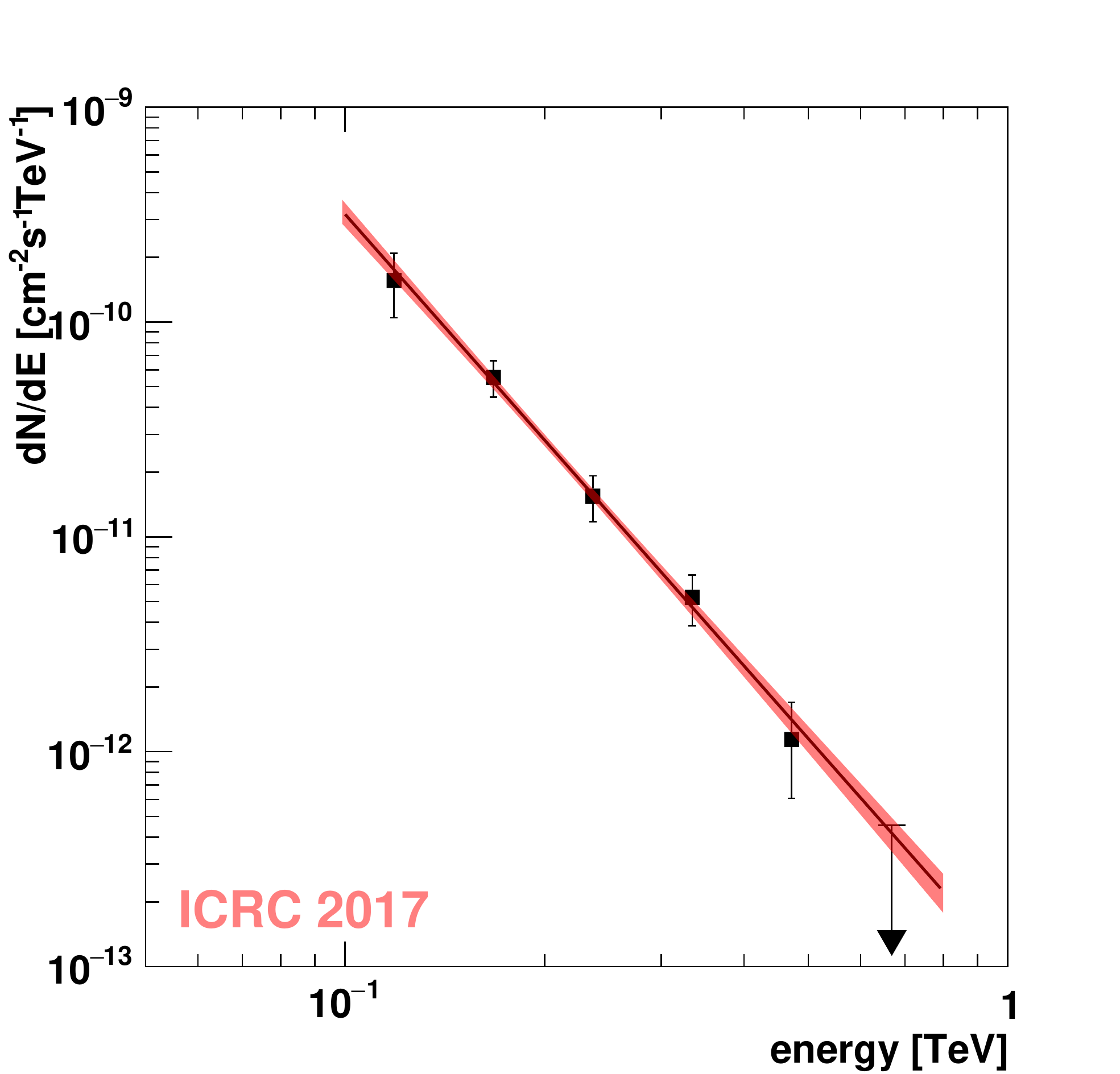}
  \caption{Differential energy spectrum of OJ~287.}
  \label{veritas:spectrum}
\end{minipage}
\end{figure*}

Figure \ref{veritas:skymap} shows the significance sky map of the vicinity of OJ~287. 
To obtain the best fit location of the VHE emission, a $\chi^2$-fit of a 2-dimensional symmetrical gaussian was fitted to the excess counts maps, 
with the $\chi^2$-statistic being minimised at the location (J2000) R.A.: (08h 54$'$ 49.1$''$) $\pm$ (2.2$''$)$_{stat}$, Dec: (+20$^\circ$ 05$'$ 58.89$''$) $\pm$ (31.96$''$)$_{stat}$. The systematic error on the VERITAS pointing accuracy is $< 25''$.

\par The time-averaged differential energy spectrum was obtained and fitted by a power-law model of the form $dN/dE = N \left(E/E_0\right)^{-\Gamma}$,
where $N$ is the flux normalisation at the normalisation energy $E_0$ and $\Gamma$ is the spectral index.  
The total time-averaged spectrum is satisfactorily fit $\left(\chi^2/NDF = 0.5/3\right)$ by a power law with parameters $\Gamma = 3.49 \pm 0.28$, $N = (2.82 \pm 0.34)\times10^{-11}~\mathrm{cm^{-2}s^{-1}TeV^{-1}} $ at the normalisation energy $E_0 = 0.2~\mathrm{TeV}$ in the energy range $100~\mathrm{GeV}-560~\mathrm{GeV}$.
The time-averaged spectrum along with the best-fit power law and $1\sigma$ confidence interval on the fit (statistical errors only) are shown in Figure \ref{veritas:spectrum}. 
Energy bins with $< 5$ on-source counts and an excess significance $< 2 \sigma$ are plotted as 95\% confidence level (c.l.) upper limits and are excluded from the fit. 
The total time-averaged flux above a threshold of 150 GeV for the entire dataset was calculated to be $(4.61 \pm 0.62)\times 10^{-12}~\mathrm{cm^{-2}s^{-1}}$ or $(1.3 \pm 0.2)$\% of the Crab Nebula flux (Crab, \cite{crab_spec}).

\par Based on the $\gamma$-ray excess observed by VERITAS, the data were divided into 3 periods of approximate constant signal, defined in Column 1 of Table \ref{veritas:period_results}.
The detection significance (shown in Column 3 of Table \ref{veritas:period_results}), indicates that only observations from period 2 result in a significant detection. 
The time-averaged flux for each period was calculated assuming a spectral index of $3.49$ and is shown in Columns 4 and 5 of Table \ref{veritas:period_results}.
Upper limits at the 95\% c.l. were calculated for the first period, which has significance $\le 2 \sigma$.
The nightly-binned flux above $150~\mathrm{GeV}$ was calculated assuming a spectral index of $3.49$ and is shown in the top panel of Figure \ref{MWL:lightcurve}, with the time-averaged flux plotted as a dashed line.
Flux points are plotted regardless of the excess significance.

\begin{table}[!htbp]
  \centering

		\resizebox{\textwidth}{!}{
			\begin{tabular}{|c|c|c|c|c|c|}
                        \hline
      Period & Exposure & Excess & $F(E>150~\mathrm{GeV})$ & $F(E>150~\mathrm{GeV})$ &$\chi^2/DOF$  \\
			& & Significance & & & \\
      (MJD) & (Hours) & ($\sigma$) & ($\times 10^{-12} [\mathrm{cm^{-2}s^{-1}}]$) & (\% Crab ) & \\
                        \hline
      (1) 57731 - 57777 & 5.0 & 2.0 & $ < 7.63$ & $< 2.1$ & $3.1/3$  \\
      (2) 57785 - 57817 & 25.3 & 10.1 & $(6.51 \pm 0.93)$ &  $(1.8 \pm 0.3)$ & $7.3/13$  \\
      (3) 57827 - 57843 & 20.1 & 2.8 & $(2.58 \pm 0.91)$ & $(0.7 \pm 0.2)$ & $20.2/13$ \\
                        \hline
      Total & 50.0 & 9.7 & $(4.61 \pm 0.62)$ & $(1.3 \pm 0.2)$ &  $44.0/31$  \\
                        \hline
                \end{tabular}}

  \caption{Summary of VERITAS observations.}
        \label{veritas:period_results}
\end{table}

\section{Multiwavelength Observations}

\label{sec:MWL}

\subsection{Fermi-LAT Observations}

The Fermi Large Area Telescope ({\it Fermi}-LAT) is a space-based pair-conversion $\gamma$-ray telescope, sensitive to $\gamma$-rays with energies from about $20~\mathrm{MeV}$ to greater then $300~\mathrm{GeV}$ \cite{FermiLAT}.
Its large field of view (2.4 sr) allows for the complete coverage of the whole sky every $\sim$3 hours.
Data from {\it Fermi}-LAT were analysed using the Science Tools package ({\it v10r0p5}) with Pass-8 ({\it P8R2}) instrument response function with the Enrico {\it Fermi}-LAT analysis suite.\footnote{\url{https://enrico.readthedocs.io/en/latest/}}
``Source''  class (evclass=128) events from both the front and back (evtype=3) between $0.1~\mathrm{GeV}$ and $300~\mathrm{GeV}$ within a $15^{\circ}$ radius of the source location were selected and a zenith cut of $90^\circ$ was applied to remove contamination from the Earth's limb. 
A binned likelihood analysis was performed on the events passing these criteria, with the spectra for sources between $3^{\circ}$ and $17^{\circ}$ away from the source frozen to their 3FGL \cite{3FGL} values and sources within $3^{\circ}$ allowed to vary.
\par The time-averaged spectrum was obtained for each of the analysis periods defined in Table \ref{veritas:period_results}. 
A power law fit is applied to each of the periods and the time-averaged flux between $0.1~\mathrm{GeV}$ and $300~\mathrm{GeV}$ is obtained.
Results of this analysis are summarised in Table \ref{fermi:period_results}. 

\par The data were binned into 5-day bins and the integral flux between $0.1~\mathrm{GeV}$ and $300~\mathrm{GeV}$ was calculated, assuming a spectral index of 2.0.
For bins with a test statistic (TS) $<$ 9, a 95\% c.l. upper limit is calculated.
The resulting light curve and shown in the middle panel of Figure \ref{MWL:lightcurve}, with the time-averaged 3FGL flux level rescaled between $0.1~\mathrm{GeV}$ and $300~\mathrm{GeV}$ (assuming the 3FGL power law fit between these energies) plotted as a dashed line. 

\begin{table}
	\begin{center}
                \resizebox{0.8\textwidth}{!}{ 
		\begin{tabular}{|c|c|c|c|}
			\hline
			Period & Test Statistic & $F(0.1~\mathrm{GeV} < E < 300~\mathrm{GeV})$ & Spectral Index \\
			(MJD) &  & ($\times 10^{-8} [\mathrm{cm^{-2}s^{-1}}$]) & \\
			\hline
			(1) 57731 - 57777 & 122.5 & $(3.97 \pm 1.07) $ & $ (1.90 \pm 0.13) $ \\
			(2) 57785 - 57817 & 100.9 & $(5.01 \pm 1.44)$ &  $(1.96 \pm 0.16)$ \\
			(3) 57827 - 57843 & 95.8 & $(4.78 \pm 1.89)$ & $(1.75 \pm 0.19)$ \\
			\hline
			3FGL & & $ (7.86 \pm 0.28)  $ & $(2.12 \pm 0.03)$ \\
			\hline
		\end{tabular}
		}
	\end{center}

	\caption{Summary of {\it Fermi}-LAT observations. The 3FGL fit is applied between $1~\mathrm{GeV}$ and $100~\mathrm{GeV}$ giving a flux of $(5.90 \pm 0.21)\times10^{-9} \mathrm{cm^{-2}s^{-1}}$; the quoted flux is obtained by assuming this power law fit is accurate between $0.1~\mathrm{GeV}$ and $300~\mathrm{GeV}$ and assuming the same percentage error.}
	\label{fermi:period_results}
\end{table}

\subsection{Swift Observations}
The \textit{Swift} satellite \cite{Swift} carries the focusing X-ray telescope (XRT), which is sensitive to photons with energies between 0.3 and 10 keV.  
OJ~287 has been observed more than 100 times since December 2016.  
The XRT event files were cleaned and calibrated with {\it xrtpipeline} using the CALDB version 20160502 calibration files. 
The first 150 seconds of all windowed-timing observations were removed in order to avoid misinterpreting the instrument's settling as variability.  
The on-source and off-source regions were circular, with radii of approximately 45$''$, encompassing approximately 80\% of the XRT point spread function (PSF). 
In all cases, the background region was defined with a similarly sized circular region placed on a region which lacked any obvious sources. 
The ancillary response files for each observation were generated with the {\it xrtmkarf} task, applying corrections for PSF losses and CCD defects using cumulative exposure maps. 
The latest response matrices (v.014) available in the \textit{Swift} CALDB were used. 
The spectral analysis was completed with \texttt{XSPEC} for dates MJD 57786 and 57804.  
The source spectra for each observation were extracted from the cleaned event files that were grouped to require a minimum of 20 spectral counts per bin, enabling the use of $\chi^2$ statistics in the fitting of an absorbed power-law model. 
The neutral hydrogen density ($N_{\rm HI}$) measured in the LAB survey of 2.49$\times10^{20}$cm$^{-2}$ \cite{kalberla} was used in the spectral fit.

\section{Discussion}
\label{sec:Discus}
\begin{figure*}[!htbp]
        \centering
		\includegraphics[width=10cm]{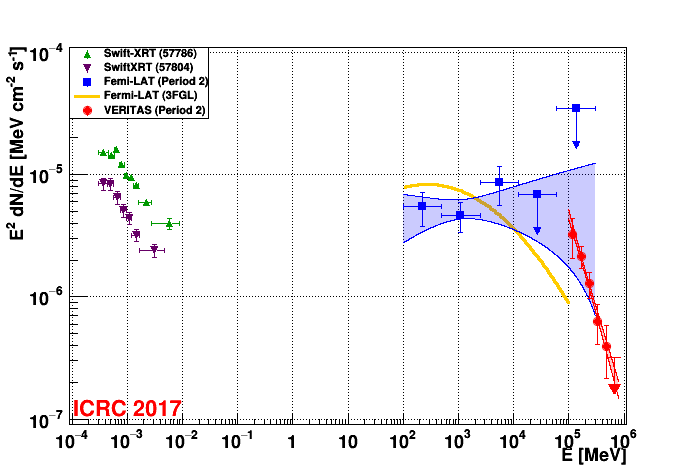}
		\caption{Spectral energy distribution for OJ~287 for period 2. For comparison the time-averaged 3FGL spectrum is plotted as an orange line. }
        	\label{MWL:comp_SED}
\end{figure*}

To explore the changing shape of the SED, the SED for period 2 is obtained and shown in Figure \ref{MWL:comp_SED}.
For period 2 the best-fit differential VHE spectrum is well fitted $\left(\chi^2/NDF = 0.25/3\right)$ by a power law with parameters $\Gamma = 3.58 \pm 0.32$, $N = (3.90 \pm 0.51)\times10^{-11}~\mathrm{cm^{-2}s^{-1}TeV^{-1}} $ at the normalisation energy $E_0 = 0.2~\mathrm{TeV}$ in the energy range $100~\mathrm{GeV}-560~\mathrm{GeV}$.
In addition the time-averaged {\it Fermi}-LAT spectrum is obtained for period 2.
The flux density is obtained for 5 equally spaced logarithmic bins between $0.1~\mathrm{GeV}$ and $300~\mathrm{GeV}$, with 95\% c.l. upper limits calculated for energy bins with a TS $<$ 9.
The VHE spectrum suggests that the peak-energy for the IC-component is located below $100~\mathrm{GeV}$. 
The {\it Fermi}-LAT spectral index for period 2 doesn't indicate a clear peak in the HE spectrum. 
The 3FGL spectrum is best-fit by a log-parabola model, indicating a IC-peak in the MeV to sub-GeV range.
To test for curvature a log-parabola model was applied to the {\it Fermi}-LAT spectrum, however no significant curvature was found.
In addition the best-fit power-law spectral index of ($1.96 \pm 0.16$) suggests that the IC-peak energy may be located in the higher-energy range detectable by {\it Fermi}-LAT where the lack of statistics effects the confidence of the fit, possibly resulting in the lack of significant spectral curvature.
This, in contrast to the 3FGL spectrum, indicates a shift in the IC-peak energy to higher energies during the period of enhanced X-ray activity, pushing the VHE emission into a level detectable by current generation IACTs.
The {\it Swift}-XRT SEDs for the hardest-brightest and the softest-dimmest nights during period 2 are also plotted in Figure \ref{MWL:comp_SED} (green and pink points respectively). The best fit X-ray spectra are fitted by power laws of index ($2.59 \pm 0.06$) (hardest-brightest) and ($2.70 \pm 0.13$) (softest-dimmest).
Full modelling of the broadband SED for each period will be discussed in a future VERITAS publication.

\begin{figure*}[!htbp]
        \centering
		\includegraphics[width=0.9\textwidth]{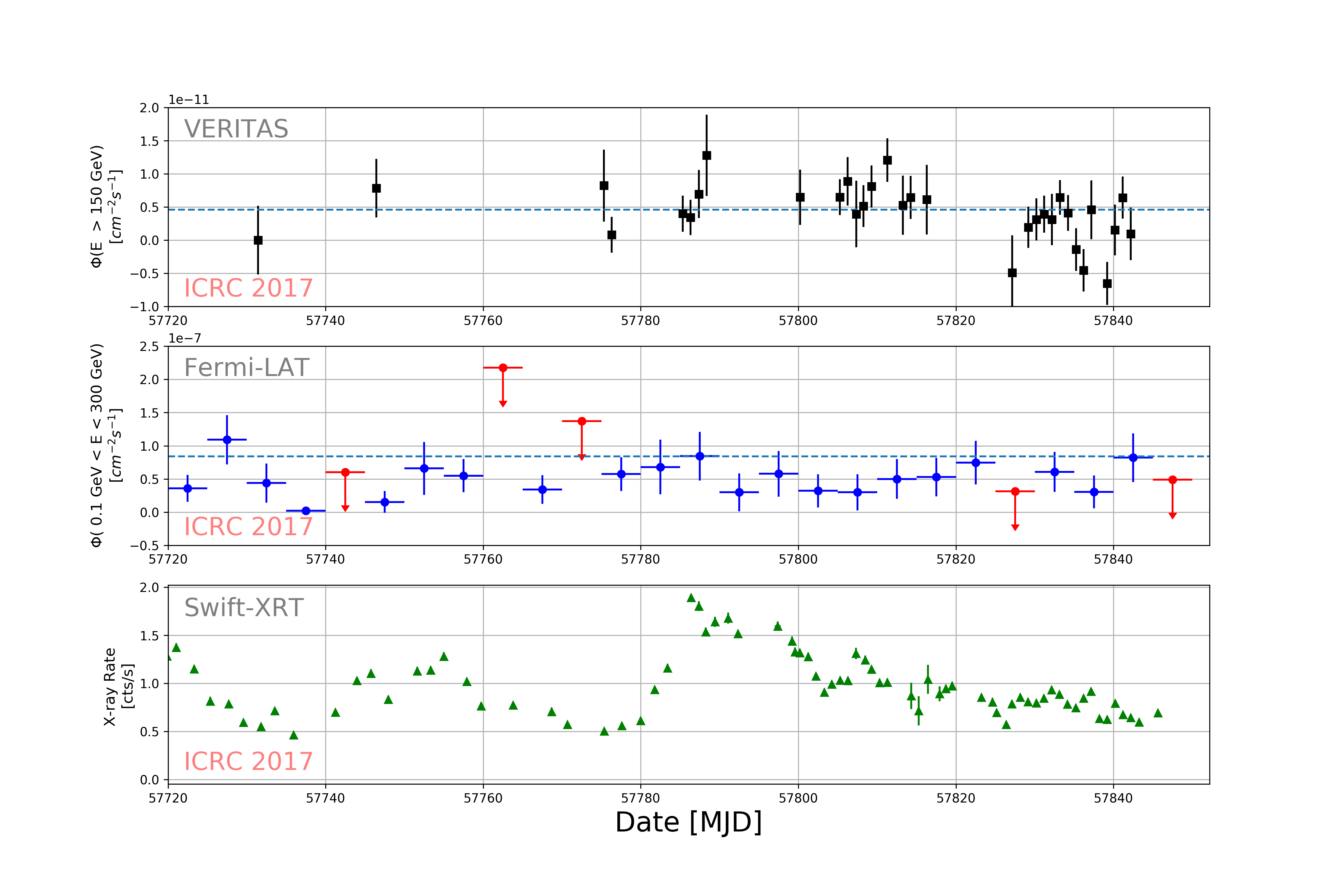}
        	\caption{Multiwavelength light curves for OJ~287. The top panel shows the nightly-binned VERITAS flux ($E > 150~GeV$). The middle shows the {\it Fermi}-LAT flux ($0.1~GeV < E < 300~GeV)$ in 5-day bins, 95\% c.l. upper limits are plotted for bins with a $TS < 9$. The bottom panel shows the nightly-binned {\it Swift}-XRT count rate. }
        	\label{MWL:lightcurve}
\end{figure*}

To test the stability of the VHE flux, a $\chi^2$-fit to a constant flux model was applied to the total nightly-binned VHE light curve and within each period.
The $\chi^2/DOF$ for the total dataset and for each period are reported in Column 6 of Table \ref{veritas:period_results}.
Periods 1 and 2 are well fit by the constant flux model with $\chi^2/DOF$ of 3.1/3 and 7.3/13 respectively.
The $\chi^2/DOF$ for period 3 and the total dataset indicate marginal disagreement with a constant flux model.
This is also shown by a $\sim$60\% decrease in the VHE flux level between periods 2 and 3.
This observed decrease in flux level is also apparent in the {\it Swift}-XRT counts rates, shown in Figure \ref{MWL:lightcurve}, which show a gradual decrease in X-ray count rates with time.
A detailed analysis of the VHE flux variability and its correlation to other wavelengths will be presented in a future VERITAS publication.

\section{Summary}
\label{sec:Summary}
VERITAS observations of OJ~287 between December 2016 and March 2017 have resulted in a strong $9.7\sigma$ detection of the source at a location consistent with the published radio location.
The time-averaged energy spectrum is best-fitted by a power-law model of spectral index $\Gamma = 3.49 \pm 0.28$. 
SED analysis of the VHE bright period shows a shift in the location of the IC-peak energy when compared with the 3FGL spectrum. 
A decrease in the VHE flux of $\sim$60\% is observed between two periods of observations; this decrease in emission is also apparent from X-ray observations taken by {\it Swift}-XRT.
A full study of the broadband emission for each period, including SED modelling and the relationship between the VHE emission and other wavelengths, will be discussed in a future VERITAS publication.

\section{Acknowledgements}
VERITAS acknowledgements can be found at: \url{https://veritas.sao.arizona.edu}\\
S. O'Brien is funded by a UCD Research Scholarship.

\bibliographystyle{}
\end{document}